\documentclass[preprint,12pt]{elsarticle}

\usepackage{amssymb}

\usepackage{color}
\usepackage{subfig}
\usepackage{amsmath}

\begin{document}

\begin{frontmatter}


\title{Influence of morphological instability on grain boundary trajectory during directional solidification}


\author[PMC,NEU]{Supriyo Ghosh}
\ead{gsupriyo2004@gmail.com}
\author[NEU]{Alain Karma}
\author[PMC]{Mathis Plapp}
\author[INSP]{Silv\`ere Akamatsu}
\author[INSP]{Sabine Bottin-Rousseau}
\author[INSP]{Gabriel Faivre}

\address[PMC]{Laboratoire de Physique de la mati\`ere condens\'ee, Ecole Polytechnique, CNRS, 91128 Palaiseau, France}
\address[NEU]{Department of Physics and Center for Interdisciplinary Research on Complex Systems, Northeastern University, Boston, MA 02115, USA}
\address[INSP]{Sorbonne Universit\'{e}, CNRS UMR 7588, Institut des NanoSciences de Paris, case courrier 840, 4 place Jussieu, 75252, Paris Cedex 05, France}

\begin{abstract}
The interplay between the diffusion-controlled dynamics of a solidification front and the trajectory of a grain boundary groove at the solid-liquid interface is studied by means of thin-sample directional solidification experiments of a transparent alloy, and by numerical simulations with the phase-field method in two dimensions. We find that low-angle grain boundaries (subboundaries) with an anisotropic interfacial free energy grow tilted at an angle $\theta_t$ with respect to the temperature gradient axis. $\theta_t$ remains essentially equal to its value imposed at equilibrium as long as the solidification velocity $V$ remains low. When $V$ increases and approaches the cellular instability threshold, $\theta_t$ decreases, and eventually vanishes when a steady-state cellular morphology forms. The absence of mobility of the subboundary in the solid is key to this transition. These findings are in good agreement with a recent linear-stability analysis of the problem.
\end{abstract}

\begin{keyword}
Solidification\sep
Morphological stability\sep
Grain boundaries\sep
Phase-field simultions\sep
In situ experiments


\end{keyword}

\end{frontmatter}

\section{Introduction}\label{sec_intro}
The solidification of melts generally yields polycrystalline solids. During the growth of a single solid phase, for example in dilute binary alloys, many different crystallites of the same phase, but with different orientations compete at the growth front and are separated by grain boundaries that intersect the solid-liquid interface~\cite{rappazbook}. The as-cast grain structure and the distribution of crystal orientations in the growing solid material are therefore determined by the coupled dynamics of solid-liquid interfaces and grain boundaries close to the growth front~\cite{corriel73,ungar84,ungar85,sabine2002}.

Competition between grains of different orientations has been thoroughly studied for growth velocities well above the onset of morphological instability, in a regime where the solid-liquid interface is strongly morphologically unstable and forms dendritic array structures (see Refs.~\cite{Tourret15,tourret2017grain} and references therein). In this dendritic regime, grain boundaries meet the solid-liquid interface deep inside the semi-solid mushy zone (i.e. far behind the dendrite tips) and have a negligible influence on grain selection, which is predominantly controlled by the growth competition of primary, secondary, and tertiary dendrite arms from neighboring grains at the leading edge of this zone \cite{Tourret15,tourret2017grain}.

In contrast, the low-velocity regime of monophase growth remains comparatively less understood. When the interface is morphologically stable or weakly unstable, forming relatively shallow cells, the dynamics of grain boundaries is tightly coupled to the dynamics of the solid-liquid interface. This regime has been recently investigated analytically under the assumption that the departure of the interface from planarity is small \cite{gabriel2013}, but the predictions of this theory still await validation, especially in view of the fact that departure from planarity becomes significant even close to the onset of morphological instability. Here, we investigate this low-velocity regime both experimentally and computationally.

We study the dynamics of subboundary grooves  by thin-sample directional solidification experiments of transparent alloys and by numerical simulations with a phase-field model. In directional solidification, the alloy sample is translated at a constant velocity $V$ towards the cold part of an experimental setup that imposes a fixed and constant thermal gradient $G$. The crystallization front is morphologically stable for velocities below the  threshold velocity $V_{MS}$ of the Mullins-Sekerka instability \cite{mullins1964}: it remains planar and perpendicular to the thermal-gradient axis. For $V\geq V_{MS}$, it undergoes a bifurcation to cellular patterns.

\begin{figure}[h]
\centering
\includegraphics[width=8cm]{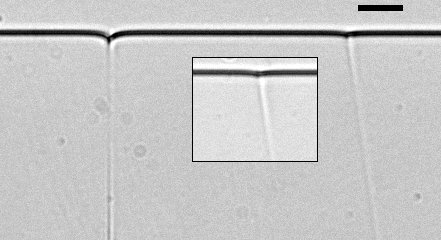}
\caption{A high-angle grain boundary (on the left) and a subboundary (on the right) intersect the solid-liquid interface at equilibrium ($V = 0$) in a thin sample of a transparent CBr$_4$-1.4~mol\%~C$_2$Cl$_6$ alloy, in a fixed temperature gradient $G$ = 120~K~cm$^{-1}$. In this image, as well as in  following ones, the thermal axis is  vertical, the liquid on the top. The solid-liquid interface appears as a thick dark line. The solid-solid interfaces are made faintly visible by thin liquid channels decorating their intersect with the sample glass walls. Note the inclination of the subboundary, and the small depth of its grain boundary groove. Inset: the same subboundary during directional solidification at low velocity ($V$~=~2.0~$\mu$m~s$^{-1}$). Bar: $20~{\rm \mu m}$.}\label{fig_gb_exp}
\end{figure}

It is  known that a grain boundary groove can provoke a pre-cellulation of the solid-liquid interface in its vicinity for velocities slightly below $V_{MS}$. This has been established theoretically for a fixed, isotropic grain boundary \cite{corriel73}. The case of a low-angle grain boundary, also called subboundary, is more complex. Subboundaries are generally strongly anisotropic, that is, their free energy depends on their inclination. Therefore, at equilibrium, a subboundary that intersects the solid-liquid interface is inclined, in the solid, in a direction that is imposed by the Young-Herring condition at the trijunction (Fig.~\ref{fig_gb_exp}). It generally makes a finite angle with the temperature gradient. What happens during solidification? We demonstrate here that there exists two very distinct regimes of subbounday growth, in qualitative agreement with the theoretical predictions of Ref.~\cite{gabriel2013}. For low velocities, the subboundary grows in a direction that is very close to its inclination at equilibrium. This implies that the triple junction moves laterally along the solidification front. In contrast, for velocities above or closely below the cellular instability threshold, the triple junction as well as the subboundary follow the direction of the temperature gradient.

Moreover, the simulations demonstrate that the grain boundary mobility plays a crucial role for the morphological evolution.
In situ observations in transparent alloys indicate that high-angle grain boundaries, that is, grain boundaries that separate two crystals with a large misorientation, are mobile, whereas subboundaries remain immobile on the micrometer-scale resolution of an optical microscope. This striking difference can most likely be attributed to the fact that high-angle grain boundaries in contact with the liquid are wet, that is, they are decorated, at equilibrium, by an atomically thin layer of disordered liquid-like material over a temperature range that extends well below the solid-liquid equilibrium (see Ref.~\cite{Asta2009} and references therein). For the very same reason, high-angle boundaries are almost isotropic, whereas subboundaries can be very anisotropic. In the numerical simulations, the grain boundary mobility can be controlled independently of the anisotropy, which makes it possible to separately study the influence of these grain boundary properties.

In this paper, we present results from in situ solidification experiments and phase-field simulations that confirm the existence of a transition in subboundary behavior, from interfacial-anisotropy-driven growth at low velocity to growth along the temperature gradient at high velocities. These results validate the main qualitative predictions of Ref. \cite{gabriel2013} relating to the role of grain boundary energy anisotropy. We also find that a mobile grain boundary behaves in a qualitatively different way than an immobile one. These results open an interesting avenue for obtaining information about grain boundary energetics and mobility from directional solidification experiments.

In the remainder of this article, we will first review the theory of Ref.~\cite{gabriel2013} in more detail and briefly summarize our experimental and numerical methods. We then present selected results of experiments and simulations that illustrate the roles of energetic and kinetic grain boundary properties on morphological evolution.

\section{Theoretical Background}\label{sec_background}
Consider directional solidification of a dilute binary alloy: the sample
is pulled with velocity $V$ in a fixed temperature gradient $G$. For thin samples,
convection in the liquid is suppressed, and the problem can be treated
in two dimensions. We use some standard approximations: equal density
of solid and liquid phases, a temperature field that is independent
of the interface shape (frozen temperature approximation), interfaces
in local equilibrium (no interface kinetics), and no solute diffusion
in the solid (one-sided model).

The stability of a planar interface was analyzed by Mullins
and Sekerka \cite{mullins1964}. Below a critical velocity $V_{MS}$, the
planar interface is stable; above this velocity, the destabilizing
effect of solute diffusion overcomes the stabilizing effects of the
temperature gradient and capillarity for perturbations in a
characteristic range of wavelengths. This velocity is usually slightly
larger than the constitutional supercooling velocity $V_{CS}$ introduced
by Tiller {\em et al.} \cite{tiller53} at which solute diffusion and
the temperature gradient balance each other. More precisely: the
diffusion and thermal lengths are given by $l_D=D/V$ and
$l_T=\Delta T/G$, respectively, where $D$ is the solute diffusivity
in the liquid, and  $\Delta T= |m|\Delta c$ the freezing range, with $\Delta c$ being the concentration
difference between the liquid and the solid at the planar interface, and $m$ the liquidus slope. Constitutional supercooling occurs
at $l_T=l_D$, which entails $V_{CS}=DG/\Delta T$. The difference between $V_{CS}$ and $V_{MS}$ increases
with the capillary length $d_0=\Gamma/\Delta T$, where $\Gamma$ is
the Gibbs-Thomson constant of the solid-liquid interface. It can be 
useful to write $V_{MS}=V_{CS}(1+\mu_c)$, where the capillary correction 
$\mu_c$ depends on $d_0$. For concentrated
alloys, $d_0$ is several orders of magnitude smaller than the other
two length scales (hence $\mu_c\ll 1$, and $V_{MS}\approx V_{CS}$), 
but for very dilute alloys, such as the one studied in the experiments described below, 
the difference between $V_{CS}$ and $V_{MS}$ can be appreciable.

\begin{figure}[h]
\centering
\includegraphics[width=7cm]{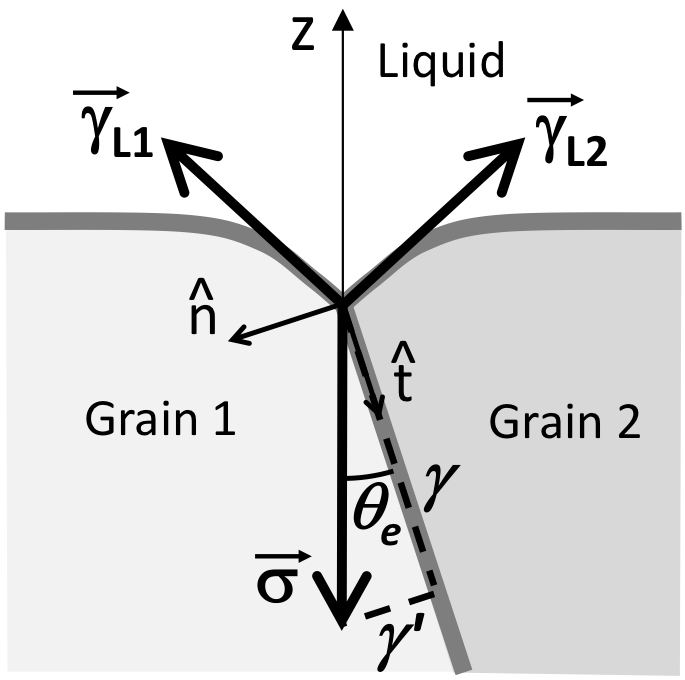}
\caption{Anisotropic force balance at a subboundary trijunction in directional solidification. Here, $\vec\gamma_{Li}=\gamma_{L}\hat{t}_{Li}$, with $i = 1,2$. See text for details.}\label{fig_gb_fb}
\end{figure}

Consider two grains, denoted by Grain $1$ and Grain $2$, that grow next
to each other into the Liquid $L$ (Fig.~\ref{fig_gb_fb}). At the intersection of the grain
boundary with the solid-liquid interface, a grain boundary groove is
formed. At the triple junction, the balance of capillary forces fixes the
dihedral angle (Fig.~\ref{fig_gb_fb}). For a thermal gradient in the $100$ K cm$^{-1}$ range, the depth of a grain boundary groove is typically of a few microns.

The influence of a symmetric grain boundary groove (that is, a shallow groove
created by an isotropic grain boundary, for isotropic solid-liquid
interfaces) on the morphological instability has been studied by
Coriell and Sekerka~\cite{corriel73}. They found that the ``built-in''
perturbation created by the grain boundary groove does not alter the
threshold velocity $V_{MS}$. However, the approach of the instability
threshold becomes visible because a stationary perturbation appears
around the grain boundary groove, which has the form of a decaying
sine wave with a spatial decay length that diverges when $V$ tends
to $V_{MS}$ from below.

This analysis was extended in Ref.~\cite{gabriel2013} to the case of
anisotropic grain boundaries. In two dimensions, the grain boundary
orientation (inclination) is specified by a single angle $\theta$.
The inclination-dependent grain boundary energy $\gamma(\theta)$ depends
on the (fixed) misorientation
between the two grains. The Young-Herring condition of capillary force
equilibrium at the trijunction point can be compactly stated using
the vector formalism developed by Cahn and Hoffman \cite{Hoffman72}.
The vector $\vec\sigma$ is given by
\begin{equation}\label{eq_sigma_vector}
\vec{\sigma} = \gamma(\theta) \hat{t} + \frac{\partial\gamma(\theta)}{\partial\theta}\hat{n},
\end{equation}
where $\hat t$ and $\hat n$ are the unit tangent and normal vectors
to the grain boundary. The Young-Herring condition is then expressed as
\begin{equation}\label{eq_yh}
\gamma_{L}\hat{t}_{L1} + \gamma_{L}\hat{t}_{L2} + \vec{\sigma} = \vec{0},
\end{equation}
where $\hat{t}_{L1}$ and $\hat{t}_{L2}$ are the tangent vectors to the
two solid-liquid interfaces at the trijunction point (see Fig.~\ref{fig_gb_fb}).
The solid-liquid interfaces are assumed to be isotropic, and their free energy is noted $\gamma_{L}$.

At rest ($V=0$), the solute concentration in the liquid is uniform, and the
problem of finding the shape of the grain boundary groove in the temperature gradient is equivalent to
the problem of the meniscus of a sessile droplet in the gravity field: for uniform solute concentration,
the Gibbs-Thomson law directly links the curvature and the temperature (here, a linear function of 
the $\mathbf{z}$ coordinate) in the same way as the Laplace law links curvature and pressure for a sessile droplet.
The groove profile is thus universal, and must in particular be mirror symmetric with respect to the 
$\mathbf{z}$ axis. As a consequence, the contact angles at the trijunction point are also the
same, and $\vec\sigma$ must be parallel to $\mathbf{z}$.
This yields a sufficient condition to determine the inclination
angle of the grain boundary at equilibrium. Expliciting the components
of $\hat t$ and $\hat n$ in terms of $\theta$, it is found that
\begin{equation}\label{eq_gb_tilt}
\gamma(\theta)\cos \theta + \frac{\partial\gamma(\theta)}{\partial\theta} \sin \theta = 0.
\end{equation}
We note $\theta_e$ the solution of this equation.

In thin-sample directional solidification experiments of transparent
alloys, grain boundaries are made visible by the presence of a thin 
liquid channel that decorates their intersect with the glass walls.
It is observed that a subboundary groove at the solid-liquid interface
presents a finite dihedral angle, and a much smaller depth than
large-angle (wet) grain boundaries. This implies that the subboundary
energy, or, rather, $|\vec\sigma|$, is smaller than twice the
solid-liquid surface free energy $\gamma_{SL}$. Since the
cross-section of the liquid channel along the boundary is likely
to decrease when $|\vec\sigma|$ decreases, a
subboundary exhibits a fainter contrast than a large-angle grain boundary in optical images.

\begin{figure}[h]
\centering
\includegraphics[width=7cm]{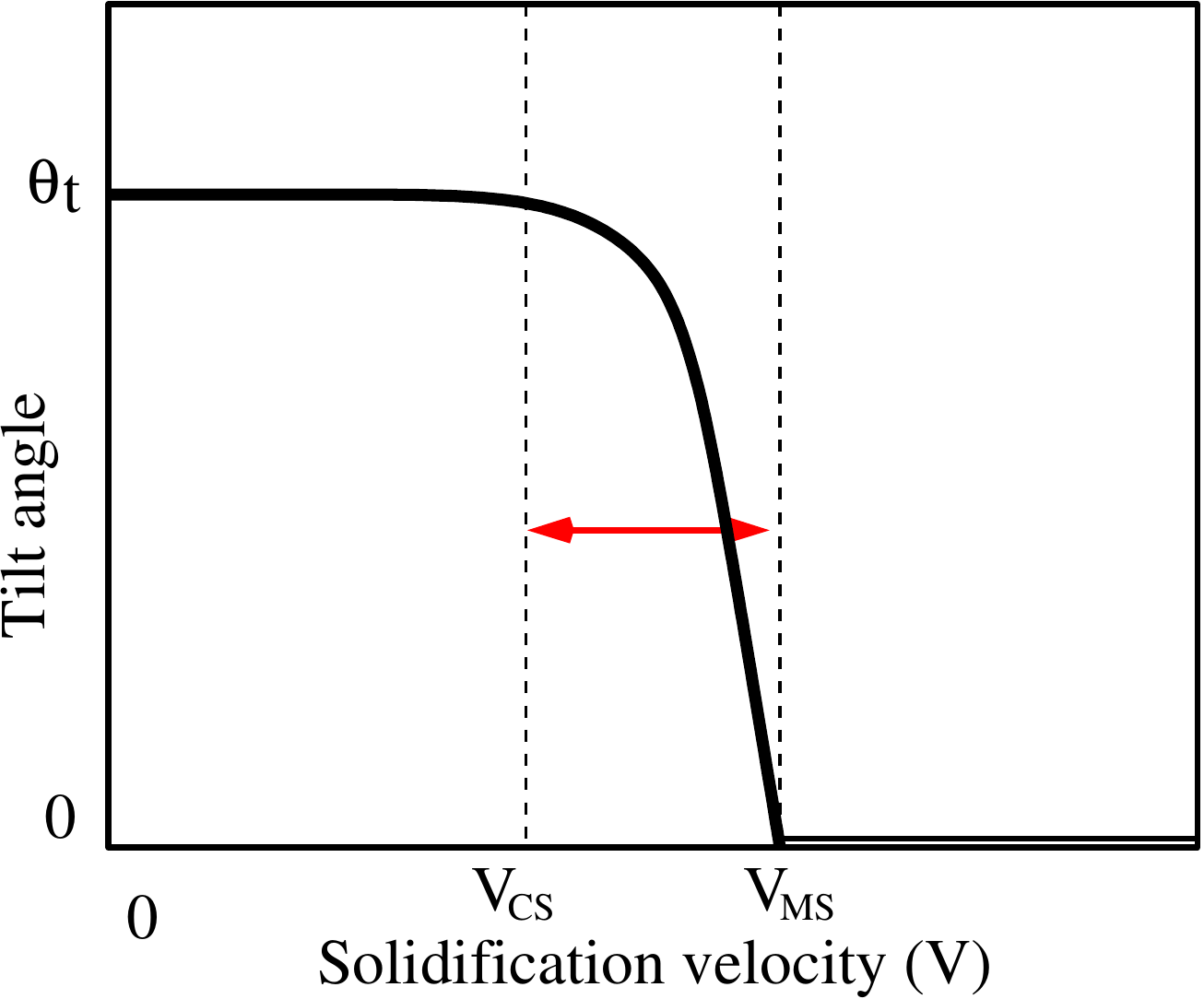}
\caption{Schematic representation of the subboundary tilt angle $\theta_t$ as a function of the solidification velocity $V$ in steady state, as found theoretically in Ref. \cite{gabriel2013}. There are three distinct regimes: (i) for $V<V_{CS}$, $\theta_t$ is essentially equal to the equilibrium inclination angle $\theta_e$; (ii) for $V_{CS}<V<V_{MS}$, $\theta_t$ decreases with $V$, and (iii) for $V>V_{MS}$, $\theta_t$ vanishes and the subboundary grows parallel to $\mathbf{z}$.} \label{fig_gb_tilt}
\end{figure}

In Ref.~\cite{gabriel2013}, the behavior of such a trijunction was
examined for non-zero $V$ in the limit of small slopes and for a
mild anisotropy, that is, a function $\gamma(\theta)$ that has no
cusps or missing orientation. Generally speaking, it is expected that the subboundary 
remains tilted at a certain angle $\theta_t$, and thus the trijunction travels sideways at a velocity $V_{t}=V\tan\theta_t$.
It was found in  Ref.~\cite{gabriel2013} that for velocities below
$V_{CS}$, the grain boundary orientation remains essentially the same
as at equilibrium ($\theta_t \approx \theta_e$; see Fig. \ref{fig_gb_fb}). For velocities
closely below the Mullins-Sekerka threshold and beyond, the tilt
angle becomes zero, and the trijunction travels along the $\mathbf{z}$
axis. The angle $\theta_t$ changes continuously between these two regimes in the velocity
range between $V_{CS}$ and $V_{MS}$, as schematically depicted in Fig.~\ref{fig_gb_tilt}.

Qualitatively, this transition can be understood in the following way. As long as the solid-liquid  interface is 
morphologically stable, it remains planar outside of the grain boundary groove, which implies that the capillary 
force balance at the trijunction (Young-Herring condition) gives a strong constraint
for the inclination of the subboundary, even though the system is not in equilibrium any more. In contrast, once
the morphological instability sets in, the interface becomes curved and the concentration becomes inhomogeneous
along the interface. Therefore, the grain boundary groove can become asymmetric, and the inclination of the
subboundary can adjust. For a strongly cellular interface, the subboundary is attached to a deep groove between
cells, and is thus completely slaved to the dynamics of the cellular solid-liquid interface.
 
It should be mentioned that similar arguments have recently been used to analyze the related problem of lamellar eutectic growth  \cite{silvere2012t}. In this case, two different solid phases grow next to each other. If the solid-solid interphase boundaries are anisotropic, the capillary force balance determines the growth directions, under the assumption that the contact angles on both sides of a lamella remain symmetric. Since there is no morphological instability in binary eutectic alloys, only the interfacial-anisotropy driven growth regime should be observable. These predictions are in good agreement with experiments \cite{silvere2012e} and  have been nicely confirmed by numerical simulations using phase-field and boundary-integral techniques \cite{supriyo2014,supriyo2017,supriyo_thesis,supriyo2015}.


\section{Methods}\label{sec_method}
\subsection{Experimental}
We performed thin-sample directional solidification (thin-DS) experiments using a transparent alloy with nonfaceted solid-liquid interfaces, namely, ${\rm CBr_4}$-${\rm C_2Cl_6}$ (liquidus slope: $|m| = 80$~K~mol$^{-1}$; partition coefficient: $K=0.75$; capillary length: $d_0=0.23~{\rm \mu m}$; solute diffusion coefficient in the liquid: $D$ = 500~$\mu$m$^2$ s$^{-1}$) \cite{Mergy93}, of nominal concentration $C_0 = 1.4 \pm 0.1$~mol\%. Details about the thin-DS method can be found in Ref.~\cite{Akamatsu98}. In brief, for solidification, the sample is pulled at a velocity $V$ ($1-10$~$\mu$m~s$^{-1}$) along  {\bf z} toward the cold part of a steady thermal gradient $G$ (120 $\pm$ 20 K cm$^{-1}$) that is established along the {\bf z} axis between two temperature regulated copper blocks separated by a 5-mm gap. Thin rectangular samples are made of two flat glass plates separated by 12-${\rm \mu m}$ thick polymer-strip spacers that fix the thickness and the lateral dimensions ($5\times 40$~mm$^2$) of the alloy film. The solid-liquid interface was observed in real-time with an optical microscope. Images were recorded with a monochrome digital camera, and transferred to a PC for further analysis. The nominal constitutional-supercooling, and Mullins-Sekerka threshold velocities for this alloy composition are $V_{CS}$ $\approx$ 15~$\mu$m s$^{-1}$  and $V_{MS}$ $\approx$ 20~$\mu$m s$^{-1}$, respectively. However, the actual cellular threshold velocity was estimated to be of about 15 $\pm$ 4 $\mu$m s$^{-1}$. This difference with the theoretical predictions can be attributed to unidentified impurities.

\begin{figure}[htbp]
\centering
{\includegraphics[width=\textwidth]{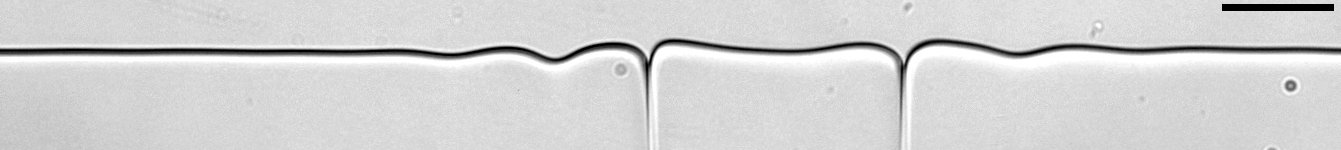}  
\caption{Pre-cellulation of the solid-liquid interface in the vicinity of two neighbouring grain boundaries at a velocity ($V$ = 9.9 $\mu$m s$^{-1}$) well below the cellular threshold velocity. Same sample as in Fig.~\ref{fig_gb_exp}. Bar: 100 $\mu m$.}\label{fig_9mu} }
\end{figure}

We therefore used some of our in situ observations for  a semi-quantitative determination of the actual value of $V_{CS}$. We observed first that for $V$ values smaller than 6 $\mu$m s$^{-1}$, the front was systematically planar over the whole width of the sample during solidification. In this regime, the depth of GB grooves was comparable to that observed at rest ($V=0$). This is shown in  Fig.~\ref{fig_gb_exp} for a subboundary. Within a finite velocity range, say, $V = 6-10$ $\mu$m s$^{-1}$, the front still remained essentially planar, except for a pre-cellulation, that is, several periods of a decaying interface oscillation, close to grain boundary grooves, as predicted by Coriell and Sekerka~\cite{corriel73}. This phenomenon is illustrated  in Fig.~\ref{fig_9mu} for two isotropic grain boundaries. As stated above, fully developed cellular patterns were systematically observed for larger solidification velocities. 

\subsection{Phase-field (PF) model}\label{sec_phase_field}
We have used a grand-canonical multi-phase-field formulation that has been developed and validated in Refs.~\cite{mathis2011, abhik2012}. In particular, this model was used in Refs.~\cite{supriyo2014,supriyo2017} to determine the growth direction of eutectic lamellae with anisotropic interphase boundaries between the two solid phases $\alpha$ and $\beta$. Here, only the main features of the model will be presented; more details can be found in Ref.~\cite{supriyo2014}.

The model uses three ($N = 3$) phase-fields, namely, $\phi_{1}$, $\phi_{2}$ and $\phi_{L}$. Those fields indicate local volume fractions of the respective phases, and thus obey the constraint $\sum\nolimits_{i=1}^{N} \phi_i = 1$. Solute diffusion in the liquid is described in terms of the diffusion potential $\mu$, which is the thermodynamic conjugate of the concentration field $c$.

We start with a grand-canonical free energy functional
\begin{equation}\label{eq_functional}
\Omega = \int \epsilon a(\phi,\vec{\nabla}\phi) + \frac{1}{\epsilon} \mathcal{W(\phi)} + \omega(\mu, T, \phi),
\end{equation}
where $\epsilon$ is a length scale parameter related to the numerical interface thickness, and 
$a(\phi,\vec{\nabla}\phi)$ is the gradient energy density given by,
\begin{equation}\label{eq_a}
a(\phi,\vec{\nabla}\phi) = \sum\nolimits_{i<j}^{N} \gamma_{ij} \left[ a_c (\hat{q}_{ij})\right]^2 |\vec{q}_{ij}|^2.
\end{equation}
Here, $\vec{q}_{ij}= \phi_{i} \vec{\nabla} \phi_{j} - \phi_{j} \vec{\nabla} \phi_{i}$ is a vector normal to the interface $i-j$,  $\hat{q}_{ij}$ is the corresponding unit vector, and $a_c$ is the anisotropy function of the $i-j$ interface.

$\mathcal{W(\phi)}$ in Eq.~(\ref{eq_functional}) is a multi-obstacle potential defined by
\begin{equation}\label{eq_w}
\mathcal{W(\phi)} = 
\begin{cases}
\frac{16}{\pi^2} \sum\limits_{i,j = 1}^{N} \gamma_{ij}\phi_{i} \phi_{j} +  \sum\limits_{i<j<k = 1}^{N} \gamma_{ijk} \phi_{i} \phi_{j} \phi_{k} & \mbox{if $\phi \in \Sigma$} \\
\infty & \mbox{elsewhere},
\end{cases}
\end{equation}
$\Sigma$ is bounded by $\phi_i \geq 0$ and $\sum\nolimits_{i}^{N} \phi_i = 1$. $\gamma_{ijk}$ is a third order potential term which avoids appearance of any unwanted ``foreign'' phases in the binary $i-j$ interfaces. 

$\omega(\mu, T, \phi)$ in Eq.~(\ref{eq_functional}) is a grand-canonical potential which is obtained by interpolating the Legendre transformations of the concentration-dependent Helmholtz free-energy densities, $\omega_i$, of the constituting phases,
\begin{eqnarray}\label{eq_gp}
\omega(\mu,T,\phi) &=& \sum\nolimits_{i=1}^{N} \omega_{i}(\mu,T) h_i(\phi) \\
\omega_{i}(\mu,T) &=& f_i - \mu c \label{eq_fe}.
\end{eqnarray}
Here $h_i(\phi)=\phi_i^2(3-2\phi_i) + 2 \, \phi_i \phi_j \phi_k$ are weight functions that satisfy $\sum_{i=1}^{N}h_i(\phi) = 1$.

The equation of motion for $\mu$ is given by
\begin{equation}\label{eq_dmudt}
\frac{\partial \mu}{\partial t} = \frac{\vec{\nabla}\cdot(M(\phi)\vec{\nabla}\mu - \vec{j}_{at}) - \sum\nolimits_{i=1}^{N} c_i\frac{\partial h_i(\phi)}{\partial t}}{\sum\nolimits_{i=1}^{N} \frac{\partial c_{i}}{\partial \mu}h_{i}(\phi)},
\end{equation}
where $c_i=-\partial\omega_i/\partial\mu$. An antitrapping current $\vec{j}_{at}$ is added to avoid artificial solute trapping effects, thereby guaranteeing that the correct thin interface limit is obtained at the solid-liquid interfaces \cite{Karma2001, Echebarria04}. The atomic mobility of the diffusing atoms $M(\phi)$ can be related to the diffusion coefficient in the liquid $D$ by $M(\phi) = \phi_l D/(\partial^2 f_i/\partial c^2_i)$.

The temporal evolution of $\phi$ follows the Allen-Cahn dynamics
\begin{equation}\label{eq_dphidt}
\frac{\partial \phi_i}{\partial t} = -\frac{1}{\tau \epsilon}\left[\frac{\delta \Omega}{\delta \phi_i}  - \Lambda \right],
\end{equation}
with $\tau$ being the relaxation time, which may vary between different interfaces according to
\begin{equation}\label{eq_tau}
\tau = \frac{\sum_{i<j}\tau_{ij}\phi_i\phi_j}{\sum_{i<j}\phi_i\phi_{j}}.
\end{equation}
The values of $\tau_{ij}$ are chosen so as to make the interface kinetics vanish \cite{abhik2012}. The Lagrange multiplier $\Lambda$ is added (in Eq.~(\ref{eq_dphidt})) to maintain the constraint $\sum\nolimits_{i=1}^N \phi_i = 1$ throughout the system.

As already mentioned, the mobility of the subboundary (solid-solid interface) plays a crucial role. Several possibilities exist to lower the mobility of the subboundary. First, one may simply choose $\tau_{12}\gg\tau_{1L}=\tau_{2L}$ in Eq.~(\ref{eq_tau}). Second, one may lower the phase-field mobility in the solid-solid interface by multiplying $\tau$ with $1-4\phi_1\phi_2$, a function that falls to zero on the interface between grains $1$ and $2$ ($\phi_1=\phi_2=1/2$). We have tested both approaches, which have yielded qualitatively similar results.

The discretized version of these equations can occasionally lead to $\phi_i$ values less than $0$ or greater than $1$. If $\phi_i$ falls below $0$, we replace it with 0, and if it increases beyond 1, we replace its value with 1. This is a standard procedure for obstacle potentials.

For the bulk free energies, $f_i$ in Eq.~(\ref{eq_fe}), we choose parabolas as described in Refs.~\cite{folch2005,supriyo2014}. As already mentioned, in the former reference we have used this model to simulate tilted lamellar eutectic growth, with three different parabolas for the three phases involved (two solids and the liquid). Here, we use two ``identical copies'' of the free energy functions for the two solids, and choose parameters for the liquid that yield a phase diagram with parallel solidus and liquidus lines, that is, $f_s=A(c-c_s^0)^2/2$ and $f_l=A(c-c_l^0)^2/2 + B(T-T_0)$ where $c_s^0$ and $c_l^0$ are the equilibrium concentrations of solid and liquid at a reference temperature $T_0$, and the constants $A$ and $B$ can be related to the liquidus slope and the latent heat of crystallization. We fix the ratio of the capillary length $d_0$ and the interface thickness parameter $\epsilon$ to $\epsilon/d_0=1.58$ in all our simulations. Unless stated otherwise, we fix the temperature gradient $G$ such as to have $l_T/d_0=100$ (in the experiments described above, $l_T/d_0\approx 200$), and vary the intensity of the cellular instability by changing $V$. The instability is predicted to occur at $l_T/l_D \approx 2.2$ ($\mu_c\approx 1.2$).

\begin{figure}[h]
\subfloat[]{\includegraphics[scale=0.55]{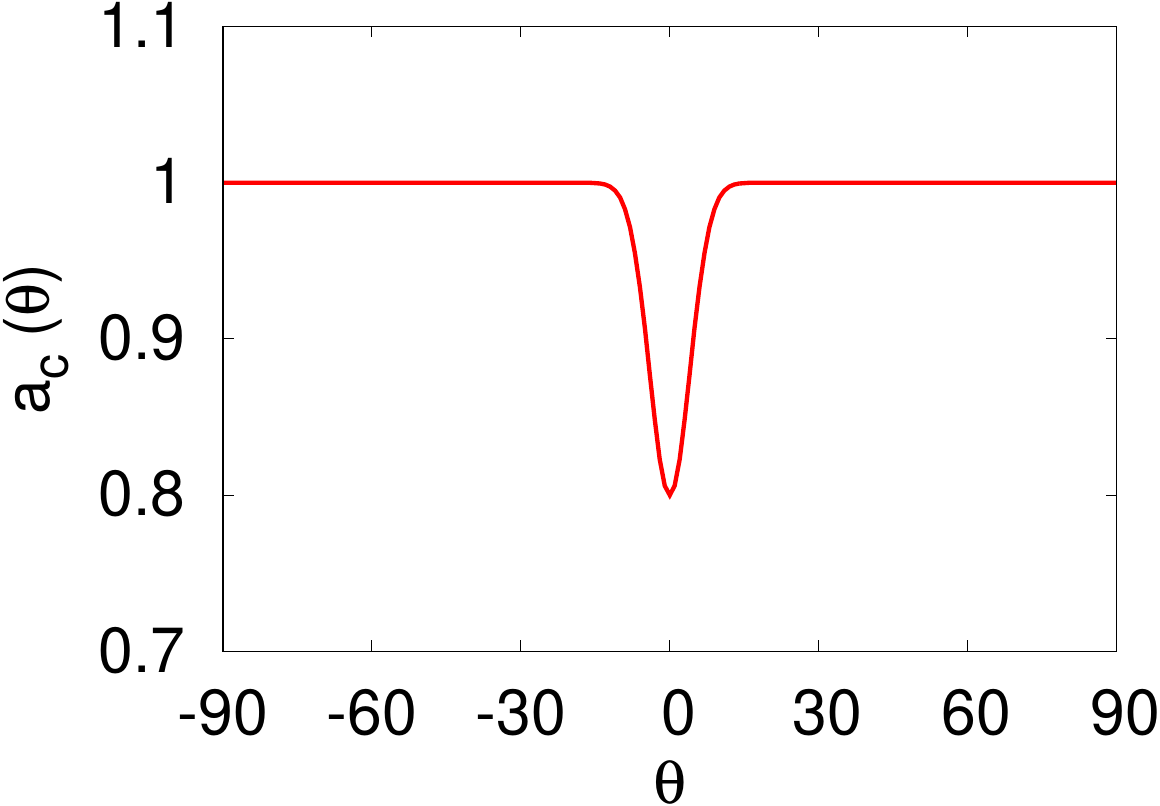}}\hfill
\subfloat[]{\includegraphics[scale=0.55]{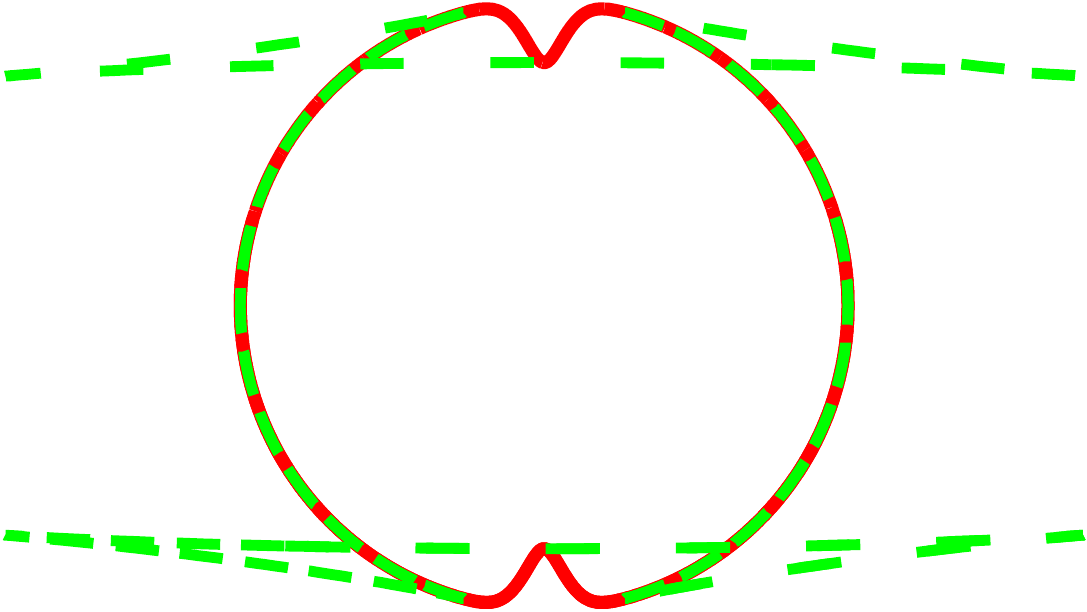}}
\caption{(a) Anisotropy function $a_c$ according to Eq.~(\ref{eq_gb_ac}), and (b) polar plot of $a_c$ (in red, full line) and the corresponding $\xi$-plot (in green, dashed line), for $\epsilon_c = 20\%$ and $w_c = 10\%$.}\label{fig_gaussian}
\end{figure}

Let us now specify the anisotropy of the boundary energy. We note $\gamma(\theta)=\gamma_0a_c(\theta)$, with $\gamma_0$ a reference value of the grain boundary energy and $a_c(\theta)$ the anisotropy function, where we have chosen $\theta = 0^{\circ}$ to correspond to a minimum in the grain boundary energy. We choose
\begin{equation}\label{eq_gb_ac}
  a_c(\theta) = 1 - \epsilon_c \left[\exp\left(-\frac{\theta^2}{w_{c}^{2}}\right)
                 +\exp\left(-\frac{(\theta-\pi)^2}{w_{c}^{2}}\right)\right].
\end{equation}
Many grain boundaries exhibit cusp-like minima in their anisotropy functions, where the energy minimum corresponds to a particular atomic configuration of high symmetry. Our anisotropy function approximates such a cusp around $\theta=0$ and $\theta=\pi$ (these two orientations are equivalent for grain boundaries in crystals whose unit cells have an inversion symmetry) with deep but smooth minima to make the grain boundary energy differentiable with respect to $\theta$. Here, $\epsilon_c$ is the magnitude of the anisotropy and $w_c$ is the width of the cusp. In all our simulations, we have chosen $\epsilon_c=0.2$ and $w_c=0.1$. The function given by Eq.~(\ref{eq_gb_ac}) is plotted for these parameters in Fig.~\ref{fig_gaussian}a, and the corresponding polar plot of the Cahn-Hoffman $\xi$-vector in Fig.~\ref{fig_gaussian}b. The equilibrium shape of a grain is given by the convex central part of the latter plot, whereas orientations on the ``ears'' are absent from the equilibrium shape. It should be noted that, by adjusting $\epsilon_c$ and $w_c$, the range of missing orientations can be tuned (and even totally avoided). Outside of the cusp, i.e. in the horizontal part of the plot in Fig.~\ref{fig_gaussian}a, the subboundary remains almost isotropic.

In order to examine the lateral drift of grain boundaries, the minimum
energy direction has to make a finite angle with respect to the temperature
gradient. This corresponds to a rigid body rotation of the entire bicrystal,
as can be physically achieved in the setup of rotating directional
solidification \cite{silvere2012e}. If the bicrystal is rotated by an
angle $\theta_R$, the grain boundary energy becomes
\begin{equation}
\gamma(\theta) = \gamma_0 a_c(\theta-\theta_R).
\end{equation}
This is implemented in the phase-field model by performing a local rotation to the crystallographic frame whenever the anisotropy needs to be evaluated, as described in Ref.~\cite{supriyo2014}.

Since the anisotropy function that we have chosen generates missing orientations, we need to apply a regularization technique in order to avoid ill-posedness of the phase-field model. As in previous studies~\cite{supriyo2014,wise2007solving,wise2005quantum}, we add the square of the Laplacian of each phase field to the free energy functional, with a prefactor $\beta$, that is, we use the free energy
\begin{equation}
\tilde\Omega = \Omega + \int_V \beta \sum_i (\vec\nabla^2 \phi_i)^2.
\end{equation} 
 The sharp corners of the equilibrium shape are smoothed out over a length scale $\sim \sqrt{\beta/(\gamma_0\epsilon)}$. Note that the variation of the extended free-energy functional generates fourth derivatives of the phase field, which imposes the use of a fine grid discretization and small time steps. 

\section{Results}\label{sec_results}
\subsection{Experiments}

\begin{figure}[htbp]
\centering
{\includegraphics[width=8cm]{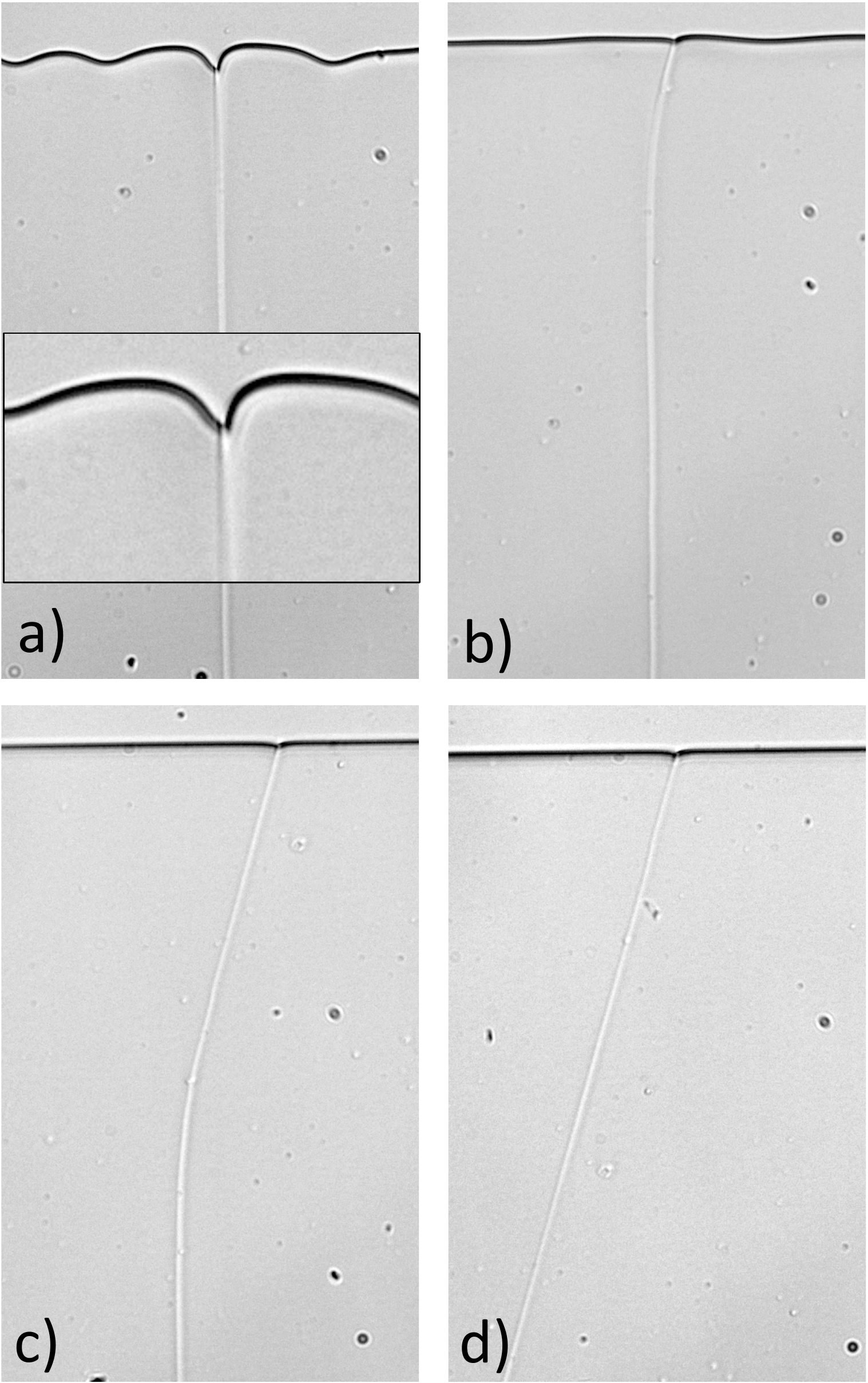} 
\caption{Dynamic behavior of a subboundary groove during directional solidification (same alloy as in Fig. \ref{fig_gb_exp}). (a) Steady state at $V$~ =~7.0~$\mu$m~s$^{-1}$ (inset: $2\times$-magnification view of the subboundary groove). (b) Beginning of the transient stage after $V$ was switched down to 3.5~$\mu$m~s$^{-1}$. (c) End of the transient  at $V$~=~3.5~$\mu$m~s$^{-1}$. d) Steady state at $V$ =~3.5~$\mu$m~s$^{-1}$. Horizontal dimension of each snapshot: 205~$\mu$m. \label{fig_7mu}} }
\end{figure}

As mentioned above, and as illustrated in Fig.~\ref{fig_gb_exp}, subboundaries exhibit two basic features:
(i) they are generally tilted with respect to the thermal gradient, and (ii) the subbounbdary groove at the solid-liquid interface presents a finite dihedral
angle, and a much smaller depth than ordinary grain boundaries.
A crucial observation here is that the trajectory of a subboundary groove, as well as its shape, remained unchanged, 
within the optical resolution, when observed at rest ($V=0$), or during directional solidification at a low rate (Fig.~\ref{fig_gb_exp}). 
There are qualitative changes when the cellular threshold is approached.
The dynamic behavior of a subboundary close to, and far below $V_{MS}$ is depicted in Fig.~\ref{fig_7mu}.
In Fig.~\ref{fig_7mu}a, the solidification conditions are similar to those of Fig.~\ref{fig_9mu},  
and one indeed observes a pre-cellulation of the solid-liquid interface, but  the subboundary groove adopts an asymmetric, 
comma-like shape, with contact angles that are different on the two sides of the boundary. This signals that the anisotropy of the subboundary remains substantial, but that the vector $\vec\sigma$ is no longer parallel to the {\bf z} axis.

Shortly after the first snapshot of  Fig.~\ref{fig_7mu}a, the velocity was lowered from $V=7.0~{\rm \mu ms^{-1}}$ down to $3.5~{\rm \mu ms^{-1}}$.
As can be seen in panel Fig.~\ref{fig_7mu}b, the perturbation of the solid-liquid interface quickly died out,
and the trijunction started to move sideways. From Figs.~\ref{fig_7mu}c and \ref{fig_7mu}d,  it is clear
that the system rapidly found a steady-state configuration with a constant lateral
drift of the subboundary trijunction, leaving a straight, but tilted subboundary in the solid. 
It can also be appreciated that the subboundary was not mobile in the solid,
since the curved section that resulted from the transient regime immediately 
after the velocity change (Fig.~\ref{fig_7mu}c) was not smoothed
out by interface motion on the time scale of the observation. A reverse transition from
inclined to straight subboundary was observed upon an increase of velocity. 

In order to test the degree of locking of subboundaries, we have also used the rotating directional solidification (RDS) method. Details about the RDS method can be found in Ref. \cite{silvere2012e}. In brief, it permits both to translate a thin sample for standard directional solidification, and to vary the orientation of the crystals by rotation with respect to the temperature-gradient axis {\bf z}.  Translation and rotation motions can be operated independently, simultaneously \cite{silvere2012e}, or in successive order \cite{INSPfaceted2009}. Here, we used the RDS method for measuring the tilt angle $\theta_t$ of a subboundary as a function of the rotation angle $\theta_R$. We used a stepwise procedure, namely: directional solidification at fixed $\theta_R$ and low velocity, measurement of $\theta_t$, partial melting, incrementation of $\theta_R$ by a few degrees, then reiteration. Using this method, we could avoid merging of the subboundary of interest with neighboring ones of opposite, or lesser inclination. The main result of this analysis is simple: $\theta_t$ varies linearly with $\theta_R$, within optical resolution, over a large  $\theta_R$ interval. Within this interval, the subboundary therefore remains locked to a fixed plane. The upper limit of the locked-state interval was observed to vary from 30 to $70^\circ$, depending on the subboundary. In most cases, that limit was observed to correspond to a sudden unlocking ($\theta_t \approx 0$) of the subboundary, and a recovery of its mobility. The unlocking process will not be discussed here.

\subsection{Simulations}

\begin{figure}[h]
\subfloat[]{\includegraphics[scale=0.5]{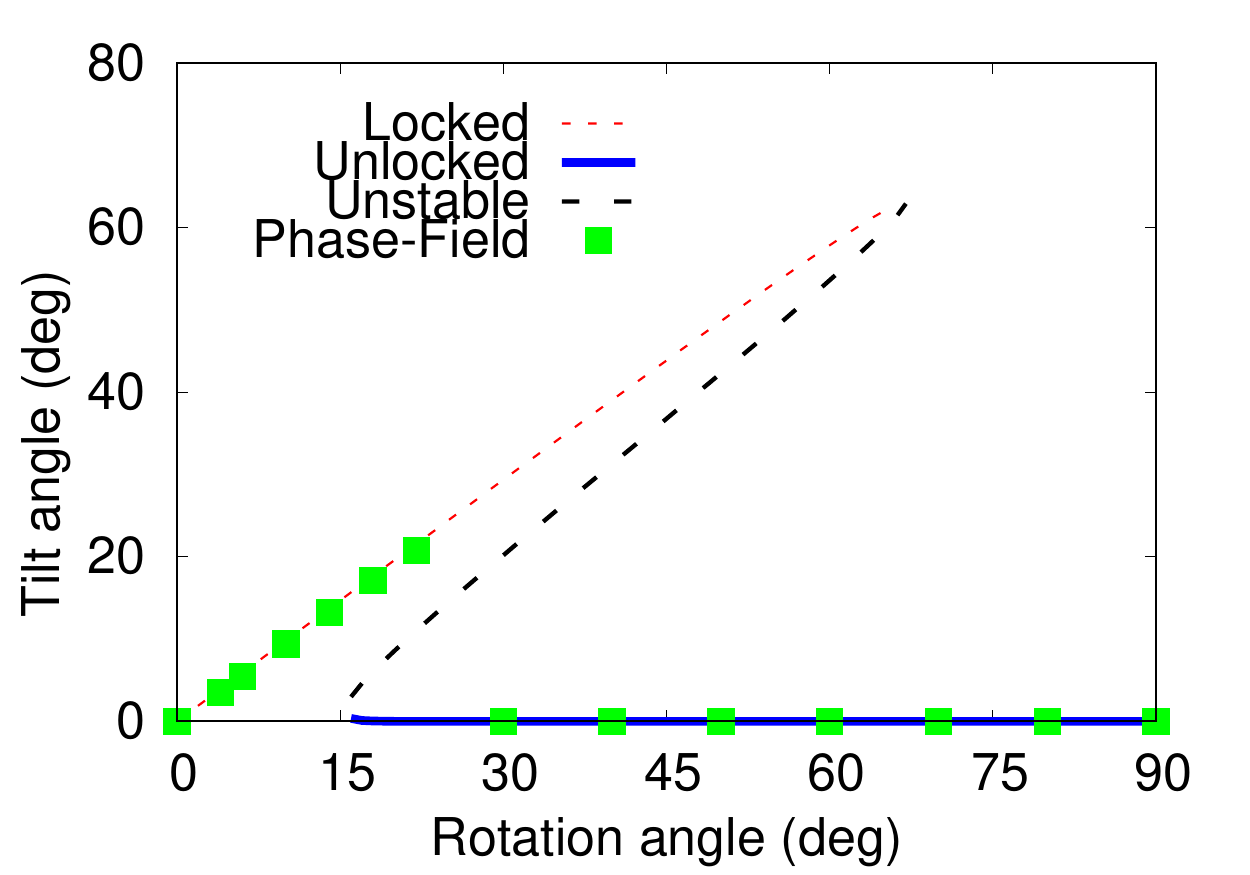}}
\subfloat[]{\includegraphics[scale=0.5]{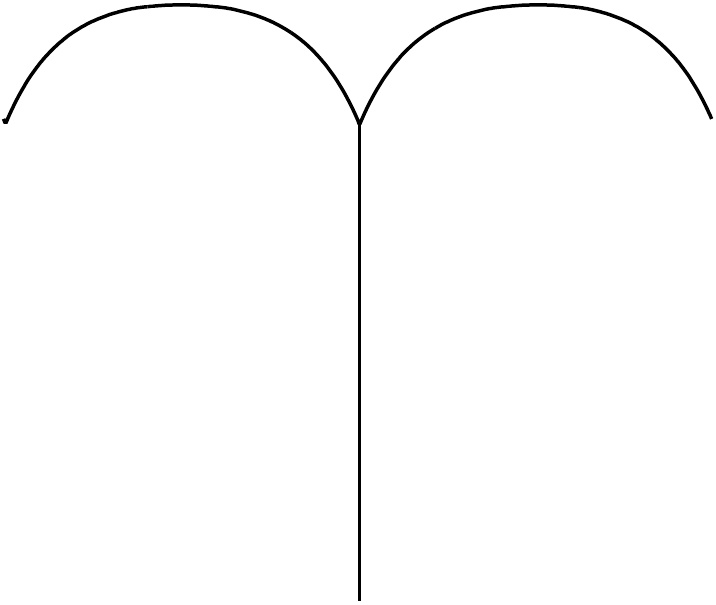}}
\subfloat[]{\includegraphics[scale=0.5]{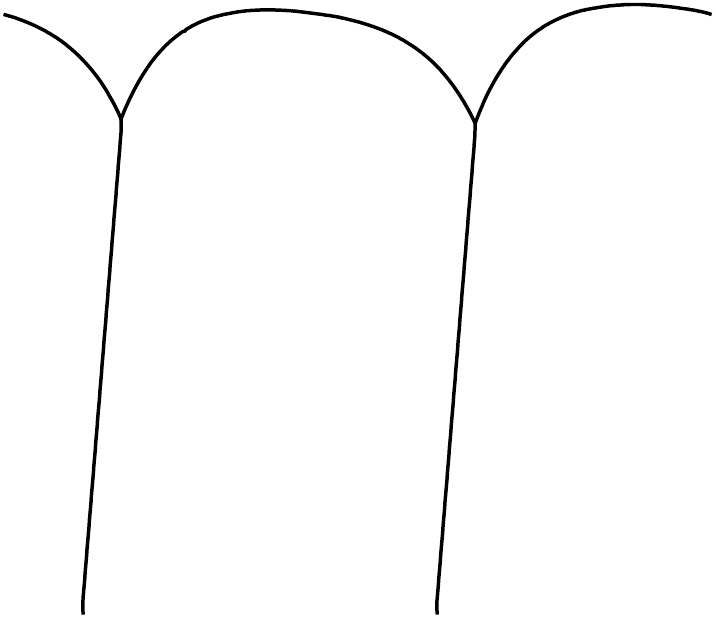}}
\caption{Below $V_{CS}$: (a) The subboundary tilt angles obtained from simulations with various rotation angles (points) are superimposed onto the analytical solutions given by Eq.~(\ref{eq_gb_tilt}) (lines). Snapshots of steady-state interface shapes are shown for (b) $\theta_R=0^\circ$, (c) $\theta_R=20^\circ$. For all simulations: $l_T/d_0 = 25$, $l_T/l_D = 0.5$, lateral system size $L_x \approx 40.5\,d_0$ (256 grid points of spacing $\Delta x=0.1\,\epsilon$).}\label{fig_bms}
\end{figure}

We have performed phase-field simulations for $V<V_{CS}$ in which the orientation of the free energy minimum was rotated with respect to the temperature gradient axis by various angles $\theta_R$. The solutions of the nonlinear Eq.~(\ref{eq_gb_tilt}) for the tilt angle are plotted for the anisotropy function used in the simulations in Fig.~\ref{fig_bms}. They can be divided into three branches, separated by the two turning points of the curve. On the  ``locked'' branch that starts at $\theta_R=0^\circ$ and extends up to the first turning point, the tilt angle essentially follows the rotation angle: the grain boundary remains nearly in the direction that corresponds to the minimum of its surface energy. On the ``unlocked'' branch beyond the second turning point, the tilt angle remains zero independently of the rotation angle: the grain boundary is now outside of the cusp in the energy, which entails that it behaves as an isotropic boundary. The solutions located on the branch that connects the two turning points are unstable.

We show two snapshot pictures of such simulations for $\theta_R=0^\circ$ (Fig.~\ref{fig_bms}b) and $\theta_R=20^\circ$ (Fig.~\ref{fig_bms}c). In all simulations, the grain boundary selects a constant steady-state inclination $\theta_t$ after a short transient, and the trijunction drifts sideways. The inclination angle is compared with the solution of Eq.~(\ref{eq_gb_tilt}) in Fig.~\ref{fig_bms}a. The simulation results fall precisely onto the ``locked'' branch between ${0}^{\circ}$ and ${20}^{\circ}$, and onto the ``unlocked'' branch between ${21}^{\circ}$ and ${90}^{\circ}$. It should be noted that Eq.~(\ref{eq_gb_tilt}) predicts the existence of two stable solutions with different tilt angles for a range of rotation angles. However, we were unable to find a value of $\theta_R$ where we could reach both solutions by our simulations. In this respect, subboundaries in this velocity regime behave similarly as anisotropic interphase boundaries in phase-field simulations of lamellar eutectics~\cite{supriyo2014,supriyo2017,silvere2015c,supriyo2015}.

\begin{figure}[h]
\begin{center}
\subfloat[]{\includegraphics[scale=0.5]{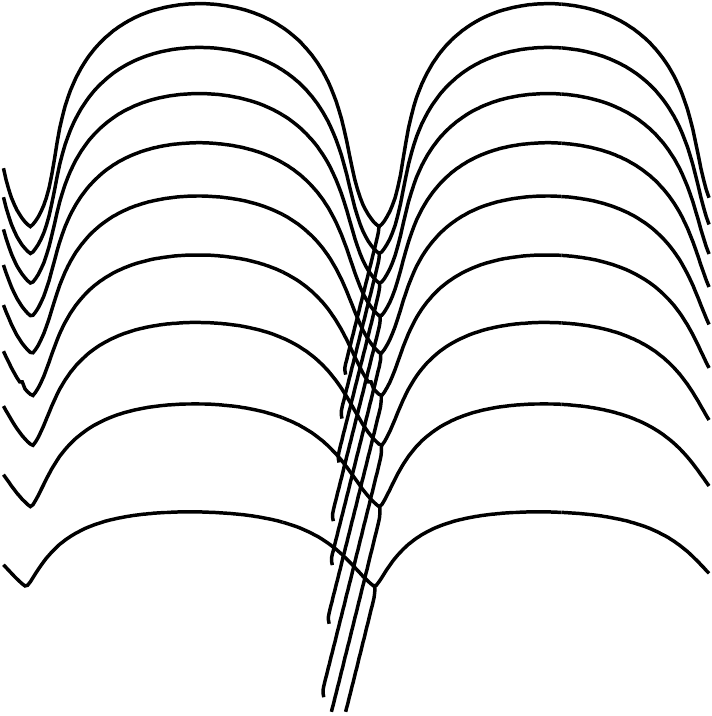}}\hspace{1cm}
\subfloat[]{\includegraphics[scale=0.5]{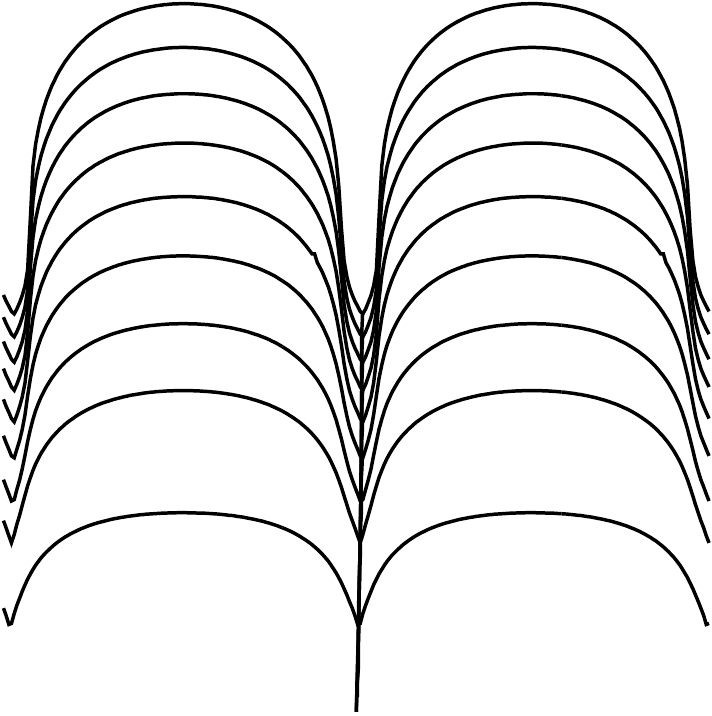}}
\caption{(a) Superimposed snapshot pictures of cell development, starting from a planar interface, far beyond $V_{MS}$ ($l_T/l_D=15.825$). (b) the same with a reduced mobility of the subboundary. Parameters: $\theta_R=20^\circ$, $l_T/d_0=100$, lateral system size $L_x \approx 40.5\,d_0$ (256 grid points of spacing $\Delta x=0.1\,\epsilon$).}\label{fig_ams}
\end{center}
\end{figure}

Above $V_{MS}$, cellular structures result with shallow or deep grooves depending on the value of the $G/V$ ratio. A simulation far beyond the instability threshold is shown in Fig.~\ref{fig_ams}a. The grain boundary still grows at a fixed inclination, which is very close to the one obtained for the same rotation angle at lower velocities. In contrast, the theory predicts that it should grow parallel to the temperature gradient in this regime. Interestingly, the trajectory of the trijunctions in the spatio-temporal plot in Fig.~\ref{fig_ams}a follows an almost vertical line, which is parallel to the growth direction. The trijunctions, therefore, obey the theoretical predictions, whereas the subboundary behind the trijunction does not.

Obviously, this behavior can only arise because the entire grain boundary moves. In fact, it is ``attached'' to the trijunction point, and since we use Neumann boundary conditions (vanishing gradients of all fields) at the lower end of the simulation box, the grain boundary can ``slide'' along the system boundary. As discussed in the section on experiments, subboundaries are usually not mobile. We have implemented two different ways in the phase-field model to lower the mobility of the solid-solid interfaces (refer to Sec.~\ref{sec_phase_field}). The results of both methods are qualitatively similar and are illustrated in Fig.~\ref{fig_ams}b. The grain boundary now just ``follows'' the trajectory of the trijunction, which results in a zero inclination angle. 

\begin{figure}[h]
\begin{center}
\includegraphics[scale=0.8]{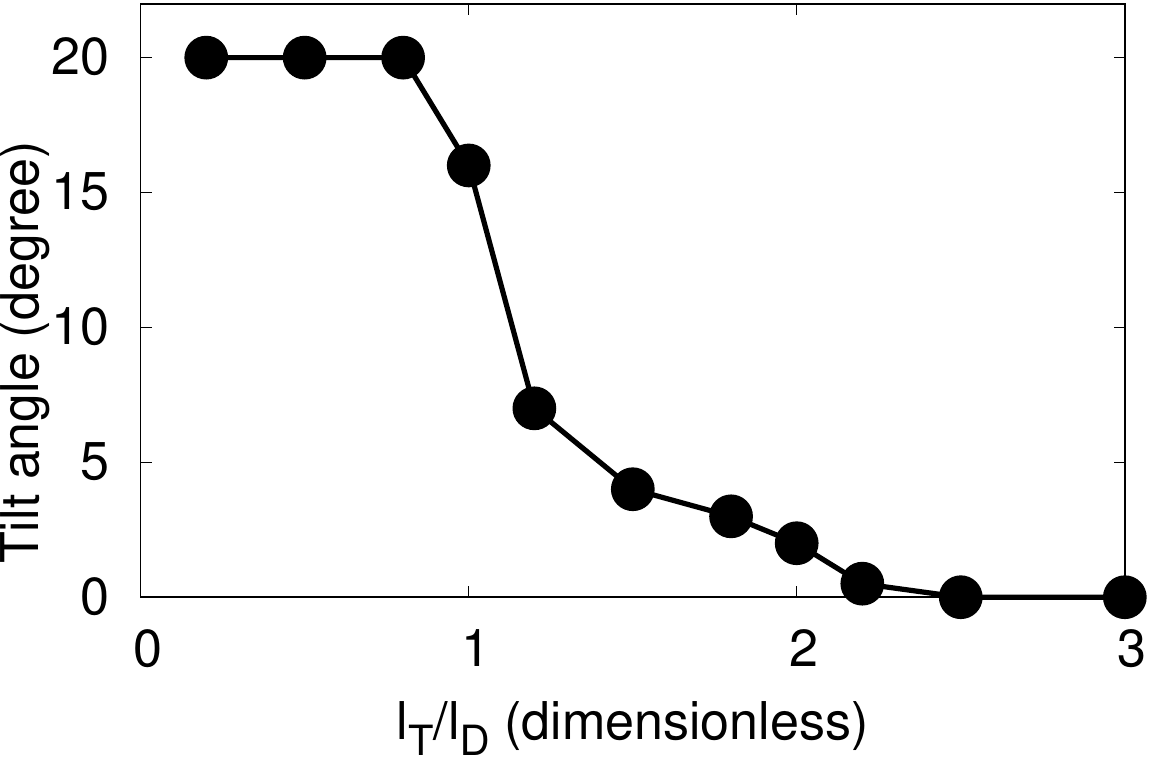}
\caption{Steady-state tilt angle $\theta_t$ as a function of the ratio $l_T/l_D$, for a fixed rotation
angle of $\theta_R=20^\circ$, $l_T/d_0=100$, and lateral system size $L_x \approx 40.5\,d_0$ (256 grid points of spacing $\Delta x=0.1\,\epsilon$).}\label{fig_tilt}
\end{center}
\end{figure}

In Fig.~\ref{fig_tilt}, we plot the steady-state inclination angle of grain boundaries with zero mobility as a function of the ratio $l_T/l_D$, which characterizes the strength of morphological instability, for a fixed rotation angle of $\theta_R=20^\circ$. The simulations largely confirm the theoretical predictions: for stable interfaces, the inclination is identical to the rotation angle (the subboundary follows its minimum energy direction), whereas for unstable interfaces, it is zero (the subboundary grows parallel to the temperature gradient). The transition between the two regimes is gradual, as sketched in Fig.~\ref{fig_gb_tilt}. This may appear surprising given that the anisotropy function that we have used in the simulations exhibits a range of forbidden orientations, from $6^\circ$ to $12^\circ$ for the chosen model parameters that control the subboundary energy anisotropy. However, a close examination of the simulation results reveals that the subboundary is actually not straight, but exhibits a zig-zag shape characteristic of unstable interfaces \cite{stewart1992spinodal,liu1993dynamics,torabi2009} when the inclination of the subboundary falls inside the forbidden range. Whereas, in this case, the local inclination of the subboundary varies, its average inclination appears to decrease smoothly with increasing growth rate as shown in Fig.~\ref{fig_tilt}. This suggests that this behavior is generic and not limited to weakly anisotropic subboundaries. The wavelength of the zig-zag pattern, however, generally depends on the regularization parameter $\beta$ of the phase-field model and is only slightly larger than the interface thickness for the parameters in the present simulations. Grain boundary dynamics could potentially become more complex in the case where this wavelength is larger than the interface thickness. This case warrants further study.

\section{Summary and Outlook}\label{sec_conclusions}

We have studied the dynamics of anisotropic low-angle grain boundaries (subboundaries) in directional solidification, both by in situ observations of thin-sample directional solidification experiments on transparent binary alloys, and by phase-field simulations. Given the fact that the anisotropy of the grain boundary energy is unknown, we did not aim for a quantitative match between experiments and modeling, but were rather interested in qualitative and generic features of subboundary dynamics at velocities close to the onset of morphological instability.

All of our results confirm the main predictions of the recent theoretical analysis presented in Ref.~\cite{gabriel2013}. At velocities that are below the constitutional supercooling threshold, the subboundary follows its minimum energy direction, which is generally inclined with respect to the axis of the temperature gradient. As a result, the triple junctions drift laterally along the growth front. At velocities that are above the onset of morphological stability, cells appear, and the subboundaries are ``slaved'' to the grooves between cells: they become aligned with the temperature gradient. The transition between the two regimes is gradual, and mostly occurs between $V_{CS}$ and $V_{MS}$. 

The simulations allow us to draw two additional conclusions, namely (i) the transition between the two regimes is gradual even for anisotropies that are strong enough to exclude certain ranges of orientations from the equilibrium shape of a grain, and (ii) the fact that the mobility of the subboundaries is low (or zero) is crucial for the transition in the inclination. Mobile subboundaries above the morphological instability are slaved to the roots of the cellular grooves, but adjust their orientation to a minimum-energy direction by moving inside the solid along their entire length.

These results open interesting perspectives for the determination of grain boundary properties. Indeed, the grain boundary inclination is a quantity that can easily be measured. If, in addition, data on crystallographic orientations can be retrieved from the experiments, as is possible in situ X-ray experiments \cite{clarke2017microstructure}, a quantitative comparison between experiments and simulations could permit to determine the anisotropy of the subboundary energy, and to gather information about their mobility. In this respect, the recently developed method of rotating directional solidification \cite{silvere2012e} is particularly valuable, since it permits to control the orientation of the sample with respect to the temperature gradient. 
This method could complement the data on grain boundary properties at high homologous temperatures, which have been obtained by other methods including molecular dynamics~\cite{yang2013solid,fensin2010structural,song2010molecular,hoyt2009method,williams2009thermodynamics} and phase field crystal~\cite{mellenthin2008phase,adland2013phase} simulations, and in particular provide a new avenue to infer the orientation dependence of grain boundary premelting from grain boundary dynamics. 

\section*{Acknowledgments}

S.G. thanks Abhik Choudhury for sharing his original phase-field code, and for many useful discussions. A.K. acknowledges support of grant DEFG02-07ER46400 of the US Department of Energy, Office of Basic Energy Sciences. This work was supported by the ANR ANPHASES project (M-era.Net: ANR-14-MERA-0004).


\end{document}